\documentclass[numberedappendix,apj,twocolumn]{emulateapj}

\pdfoutput=1

\shorttitle{The Hubble XDF}
\shortauthors{Illingworth et al.}

\begin{document}

\title{The HST eXtreme Deep Field XDF: \\Combining all ACS and
WFC3/IR Data on the HUDF Region into the Deepest Field Ever
\altaffilmark{1}}

\altaffiltext{1}{Based on data obtained with the \textit{Hubble Space Telescope} operated by AURA, Inc. for NASA under contract NAS5-26555. Data available from the Mikulski Archive for Space Telescopes (MAST) at \url{http://archive.stsci.edu/prepds/xdf/}}

\author{G. D. Illingworth\altaffilmark{2},  
D. Magee\altaffilmark{2},
P. A. Oesch\altaffilmark{2,\dag},
R. J. Bouwens\altaffilmark{3}, 
I. Labb\'{e}\altaffilmark{3}, 
M. Stiavelli\altaffilmark{4},\\
P. G. van Dokkum\altaffilmark{5},
M. Franx\altaffilmark{3}, 
M. Trenti\altaffilmark{6},
C. M. Carollo\altaffilmark{7}, 
V. Gonzalez\altaffilmark{8}
}

\altaffiltext{2}{UCO/Lick Observatory, University of California, Santa Cruz, CA 95064; gdi@ucolick.org}
\altaffiltext{3}{Leiden Observatory, Leiden University, NL-2300 RA Leiden, Netherlands}
\altaffiltext{4}{Space Telescope Science Institute, 3700 San Martin Drive, Baltimore, MD 21218, USA}
\altaffiltext{5}{Department of Astronomy, Yale University, New Haven, CT 06520}
\altaffiltext{6}{Institute of Astronomy, University of Cambridge, Madingley Road, Cambridge CB3 0HA, UK}
\altaffiltext{7}{Institute for Astronomy, ETH Zurich, 8092 Zurich, Switzerland}
\altaffiltext{8}{University of California, Riverside, 900 University Ave, Riverside, CA 92507, USA}

\altaffiltext{\dag}{Hubble Fellow}

\begin{abstract} 
The eXtreme Deep Field (XDF) combines data from ten years
of observations with the HST Advanced Camera for Surveys (ACS) and the Wide-Field Camera 3 Infra-Red (WFC3/IR) into the deepest image of the sky ever in the optical/near-IR.  Since the initial observations on the Hubble Ultra-Deep Field (HUDF) in 2003, numerous surveys and programs, including supernova followup, HUDF09, CANDELS, and HUDF12 have contributed additional imaging data across this region. Yet these have never been combined and made available as one complete
ultra-deep image dataset. We do so now with the eXtreme Deep
Field (XDF) program. Our new and improved processing techniques 
provide higher quality reductions of the
total dataset. All WFC3/IR and optical ACS data sets have been fully combined and accurately matched, 
resulting in the deepest imaging ever taken at these wavelengths ranging from 29.1 to 30.3 AB mag ($5\sigma$ in a $0.35''$ diameter aperture) in 9 filters. The combined image therefore reaches to 31.2 AB mag $5\sigma$ (32.9 at $1\sigma$) for a flat $f_{\nu}$ source.
The gains in the optical for the 4
filters done in the original ACS HUDF correspond to a typical
improvement of 0.15 mag, with gains of 0.25 mag in the deepest
areas. Such gains are equivalent to adding
$\sim$130 to $\sim$240 orbits of ACS data to the HUDF. Improved
processing alone results in a typical gain of $\sim$0.1 mag.  Our $5\sigma$ (optical+near-IR) SExtractor catalogs reveal about 14140 sources in the full field and about 7121 galaxies in the deepest part of the XDF.

\end{abstract}


\section{Introduction}

Since the first Hubble Deep Field was observed in 1995 \citep{Williams96},
the Hubble Space Telescope has demonstrated its ability to probe ever
deeper limits as new capabilities are added.  The combination of deeper
images and new cameras with sensitivity at redder wavelengths has pushed
the redshift limits well into the reionization epoch in the first Gyr of
the universe.

The Hubble UltraDeep Field \citep[HUDF;][]{Beckwith06} was observed in 2003
with the then-new Hubble Advanced Camera (ACS) that had been placed into
Hubble in 2002 by the Shuttle astronauts during the servicing mission SM3B.
ACS provided over an order of magnitude gain in discovery efficiency (the
product of survey area and efficiency) over the WFPC2 and an added redder
filter (F850LP) that opened up the universe at $z\sim5-6$
\citep[e.g.][]{Bouwens04,Bunker04,Yan04}. 

The ACS HUDF was released as a public image in 2004. The HUDF, along with
the companion wide-field GOODS images \citep{Giavalisco04a}, revealed large
numbers of $z\sim6$ galaxies \citep{Bouwens06,Bouwens07} at the end of the reionization epoch for the
first time. The NICMOS near-IR camera provided the first glimpse of galaxies at
$z\sim7$ \citep{Bouwens04Z,Yan04}, just 800 Myr after the Big Bang \citep{Thompson05}, but it was not until the new
infrared Wide Field Camera 3 (WFC3/IR) was added to Hubble in 2009 by the
Shuttle astronauts in the final Hubble servicing mission (SM4) that the
reionization epoch was opened up to extensive exploration \citep[e.g.][]{Oesch10,Bouwens10,McLure10,Bunker10,Finkelstein10}.
The remarkable gain in discovery
efficiency (over 40$\times$ for WFC3/IR compared to NICMOS) hinted at the
gains to be seen when JWST becomes operational.

\begin{figure*}[tbp]
	\centering
	\includegraphics[width=0.8\linewidth]{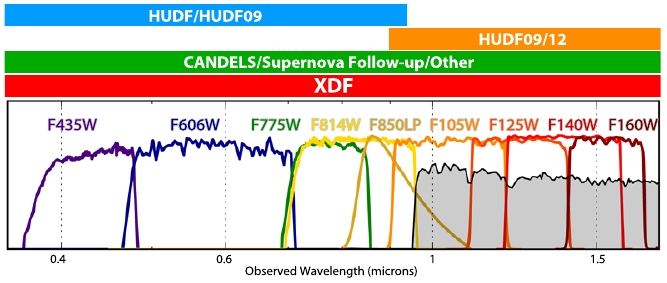}
  \caption{The XDF input datasets and spectral elements. The XDF combines \textit{all} 
the data that have been taken on the HUDF field between 2002 and 2013 - this means HUDF, HUDF09, CANDELS, HUDF12 
and many other programs. No other release has done this. This makes the deepest image from HST ever, with 
the widest spectral coverage image from HST on the HUDF. An example galaxy spectrum at $z=6.5$ is shown. 
This example galaxy is seen about 12.8 billion years ago, just 800 million years after the Big Bang. 
Due to the high fraction of neutral hydrogen present at this early cosmic time, the source is 
essentially completely invisible at optical wavelengths probed by the ACS camera and it can only be 
seen in the near-infrared with the new WFC3/IR. The wide wavelength coverage of the XDF images allow for 
the identification of similar galaxies across a very large redshift range, from $z\sim4$ to $z\sim12$.}
	\label{fig:filtersed}
\end{figure*}

\begin{deluxetable}{ll}
\tablecaption{XDF Summary \label{tab:xdfsummary}}
\tablewidth{0 pt}
\tablecolumns{2}
\startdata
\hline & \\[-3pt]
Position (J2000) &	R.A. 03h 32m 38.5s, Dec. $-$27$^\circ$~47\arcmin ~00\arcsec \\
Area (XDF total)	       &  10.8 arcmin$^2$ \\
Area	(XDF deep IR)       &  4.7 arcmin$^2$ \\
Instruments &  	ACS/WFC and WFC3/IR \\
Exposure Dates & 	July 2002 to December 2012\tablenotemark{*} \\
Total Exposure Time & 	21.7 days ($\sim$2 million seconds) \\
Number of Exposures  & 	2963 (1972 ACS/WFC; 991 WFC3/IR) \\
Typical Depths  &   $\sim$30 AB mag ($5\sigma$) in most filters \\
Combined Depth  &   31.2 AB mag ($5\sigma$) for a flat $f_{\nu}$ source   \\
Archive Link   & \url{http://archive.stsci.edu/prepds/xdf/}

\enddata
\tablecomments{Depths are for $5\sigma$ detections measured in circular apertures of 0\farcs35 diameter and are not corrected to total magnitudes.}
\tablenotetext{*}{Date when last HUDF12 exposure became publicly available on the HST archive.}
\end{deluxetable}

The HUDF was observed as part of the HUDF09 program in 2009 \citep[PI: Illingworth; e.g.][]{Bouwens11c} as one of the
first set of images taken by the new infrared camera of WFC3.
These first observations in mid-2009 were combined and released as the HUDF09 image in early 2010.
The HUDF09 observations were finally completed in the spring of 2011. Thanks to the ultra-deep coverage
in the near-infrared, these data allowed for the first systematic exploration of galaxies at $z\sim10$, resulting in the identification of then the most
distant, earliest galaxy candidate ever seen, UDFj-39546284\footnote{Although see now the discussion in \citet{Ellis13}, \citet{Bouwens13}, and \citet{Brammer13}.} 
\citep{Bouwens11a,Oesch12}.

The next step in the exploration was with the CANDELS program
\citep[PI:Faber/Ferguson;][]{Grogin11,Koekemoer11} that played a role
akin to GOODS but now with the WFC3/IR camera, providing wide-field
data that improved the statistics at higher luminosities.  These data
are an invaluable resource and contribute to the HUDF region, but the
most important addition in the near-IR was from the HUDF12
\citep[PI:Ellis;][]{Ellis13} dataset which provided additional near-IR
data from WFC3/IR over the HUDF09 field.  This dataset provided
comparable data to the HUDF09 program in some of the same filters, as
well as observations using the F140W filter.  Together the HUDF09,
CANDELS and HUDF12 datasets provide near-IR depths over part of the
HUDF area that are roughly comparable (AB mag) to that from the
original ACS dataset.

Such deep imaging in several filters across the optical to the
near-infrared is required to identify and study galaxies in the early
universe from $z\sim4$ out to $z\sim10$. Such galaxies can be selected
quite robustly due to the strong inter-galactic absorption from
neutral hydrogen \citep[e.g.][]{Madau95}. The wavelength coverage and
filter curves of the HST dataset over the HUDF, along with an example
SED for a high redshift galaxy, is shown in Figure
\ref{fig:filtersed}.

The concept of the eXtreme Deep Field (XDF) resulted from the realization
in late 2011 that all the data taken over the last ten years with the
Advanced Camera and the Wide Field Camera 3 on the Hubble UltraDeep Field
had not been combined into a single extremely deep image.  The HUDF,
HUDF09, CANDELS and numerous other datasets had been released individually,
but a combined image of \textbf{all} the images ever taken on the HUDF had
not been done.

\begin{figure*}[tbp]
	\centering
	\includegraphics[width=0.6\linewidth]{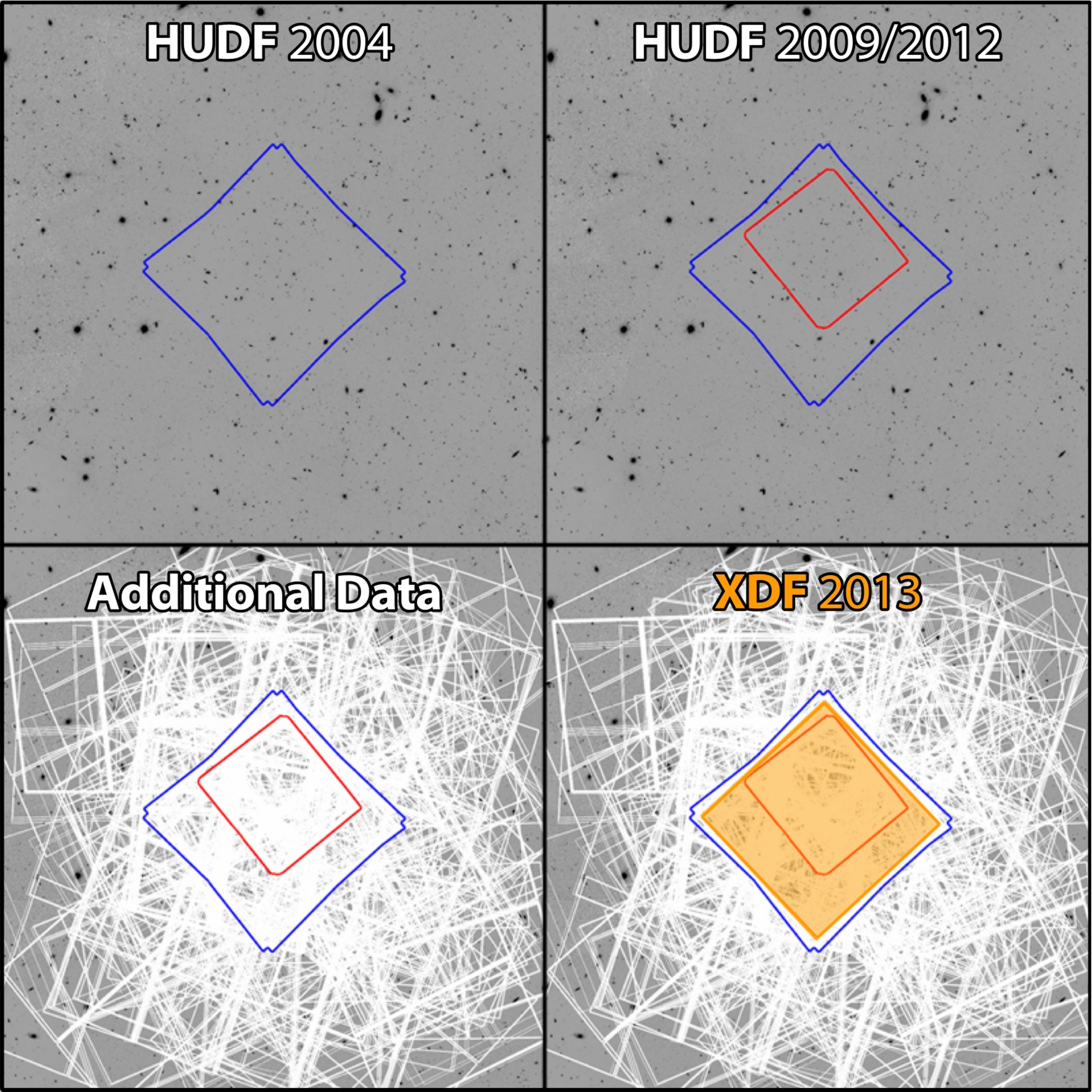}
  \caption{The steps involved in building up the eXtreme Deep Field (XDF).  The first image (top left) of the sequence shows 
the original HUDF data from the ACS. By fraction of total exposure time the original HUDF contributes 
over 50\% to the XDF, but only contains data in the optical region of the spectrum in four filters. 
In 2009, the HUDF09 project took images towards the red end of the spectrum in the near-infrared with the 
new WFC3/IR camera. These new data doubled the waveband coverage and enabled exploration for high redshift 
galaxies in the reionization epoch for the first time. The second image (top right) of the sequence adds the 
HUDF09 field with WFC3/IR that contributed around a 20\% of the data (by exposure time) to the XDF. 
Finally, the XDF/HUDF09 team took \textit{all} the other data on this region taken by numerous programs 
and combined it through a very laborious series of steps into the incredibly deep XDF image. 
As the third image (bottom left) shows, the data fall at many locations and orientations 
and much careful checking was needed to make sure all the Hubble ACS and WFC3/IR images were properly 
aligned and combined. The contributions came from seventeen other programs that comprised about 30\% of the time, 
of which the largest were the CANDELS dataset from ACS and WFC3/IR and the HUDF12 WFC3/IR dataset. 
The various supernova followup programs also made substantial contributions.
Finally, in the fourth image (bottom right) of the sequence the XDF is shown as the orange region where the 
contribution of every image over the last decade from ACS and WFC3 has been included.}
	\label{fig:buildup}
\end{figure*}

With the release of the HUDF12 observations at the end of its
proprietary period in Dec 2012 the full data set was available from
the nineteen Hubble programs that had taken observations in this
region over the past decade.  This data set was dominated by the
original HUDF, the HUDF09 and the more recent HUDF12 data, but with
important and significant contributions from many other programs,
including various supernovae followup programs and CANDELS (see
Section \ref{sec:obs} and Table \ref{tab:programs}).  In this paper,
we describe how all these datasets were combined to make the deepest
image of the sky ever.  The broad process of building up the XDF is
shown schematically in Figure \ref{fig:buildup}.

The XDF will remain the deepest image ever taken, with only marginal future
gains practical, until JWST flies.  And even then, while JWST will push to
fainter limits at $\lambda \gtrsim 0.7 \mu$m, the shorter wavelength
datasets  (F435W and probably F606W) will remain unique until a new
telescope flies that has blue optical imaging capabilities.

This paper is structured as follows: Section 2 describes the observations
that contribute to the XDF, Section 3 describes the data processing in some
detail for the ACS datasets and the WFC3/IR datasets, while Section 3.5
summarizes the data products that have been supplied for community use to the
Mikulski Archive for Space Telescopes (MAST). The tests on the datasets to verify 
their veracity are given in
Section 4, and a summary is provided in Section 5.

\section{The Observations}
\label{sec:obs}

\subsection{ACS/WFC and WFC3/IR Data over the XDF}
\label{sec:xdfdata}

Here we briefly discuss the \textit{HST} observations that were used
to create the \textit{XDF} dataset. In its current form the
\textit{XDF} includes over 10 years of ACS/WFC and 3 years of WFC3/IR
observations taken from mid-2002 through the end of 2012. These
observations include data from the original ACS optical HUDF program
(\textit{HST} PID 9978) and the WFC3/IR HUDF09 and HUDF12 programs
(\textit{HST} PID 11563 and 12498). A complete list of \textit{HST}
programs that contribute to the \textit{XDF} dataset is given in Table
\ref{tab:programs}.

\begin{deluxetable*}{ccl}
\tablecaption{HST programs contributing to the XDF\label{tab:programs}}
\tablewidth{0 pt}
\tablecolumns{3}
\tablehead{\colhead{Program ID}	& \colhead{HST Cycle} &	\colhead{Program Title}}
\startdata
9352	& 11	& The Deceleration Test from Treasury Type Ia Supernovae at Redshifts 1.2 to 1.6\\
9425	& 11	& The Great Observatories Origins Deep Survey: Imaging with ACS (GOODS)\\
9488	& 11	& Cosmic Shear - with ACS Pure Parallel Observations\\
9575	& 11	& ACS Default (Archival) Pure Parallel Program\\
9793	& 12	& The Grism-ACS Program for Extragalactic Science (GRAPES)\\
9978	& 12	& The Ultra Deep Field with ACS (HUDF)\\
10086	& 12	& The Ultra Deep Field with ACS (HUDF)\\
10189	& 13	& Probing Acceleration Now with Supernovae (PANS)\\
10258	& 13	& Tracing the Emergence of the Hubble Sequence Among the Most Luminous and Massive Galaxies\\
10340	& 13	& Probing Acceleration Now with Supernovae (PANS)\\
10530	& 14	& Probing Evolution And Reionization Spectroscopically (PEARS)\\
11359	& 17	& Panchromatic WFC3 survey of galaxies at intermediate z: Early Release Science program for Wide Field\\
& & Camera 3 (ERS)\\
11563	& 17	& Galaxies at $z\sim7-10$ in the Reionization Epoch: Luminosity Functions to $<$0.2L* from Deep IR Imaging of\\
& & the HUDF and HUDF05 Fields (HUDF09)\\
12060	& 18	& Cosmic Assembly Near-IR Deep Extragalactic Legacy Survey — GOODS-South Field, Non-SNe-Searched\\
& & Visits (CANDELS)\\
12061	& 18	& Cosmic Assembly Near-IR Deep Extragalactic Legacy Survey — GOODS-South Field, Early Visits of SNe\\
& & Search (CANDELS)\\
12062	& 18	& Galaxy Assembly and the Evolution of Structure over the First Third of Cosmic Time - III (CANDELS)\\
12099	& 18	& Supernova Follow-up for MCT (CANDELS)\\
12177	& 18	& 3D-HST: A Spectroscopic Galaxy Evolution Treasury (3DHST)\\
12498	& 19	& Did Galaxies Reionize the Universe? (HUDF12)
\enddata
\end{deluxetable*}

In order to specifically determine which \textit{HST} observations to
include in the \textit{XDF} dataset we executed searches using the MAST
\textit{HST} archive. We constrained our search to include only
\textit{HST} ACS/WFC and WFC3/IR imaging observations within a 13
arc-minute radius of the original HUDF coordinates ($\alpha=$~03:32:39.0,
$\delta=$~$-$27:47:29.0) while limiting the search to the \textit{HST}
filters F435W, F606W, F775W, F814W, F850LP, F105W, F125W, F140W, F160W, and
the exposure time to $>100$ seconds. Although, there have been many other
HST observations taken with legacy instruments (e.g WFPC2 and NICMOS) we
chose to use only imaging data from \textit{HST}'s current optical and
near-IR instruments and only filters where a substantial investment of HST
orbits had been contributed. Once we had determined the set of observations
to include in the \textit{XDF} dataset, we used the \textit{HST} MAST
archive or the Canadian Astronomy Data Centre (CADC) \textit{HST} archive
to acquire these data.

Our search resulted in a total of 2963 $HST$ exposures, 1972 in the optical filters
of ACS/WFC and 991 exposures in the NIR with WFC3/IR. These images sum up
to a total exposure time of almost 2Ms. This corresponds to about 21.7 days
of actual open-shutter observing time (13.6 days in ACS and 8.1 days in
WFC3/IR).  To acquire this amount of data would have taken $HST$ about 790
orbits or about 52 days of clock time. 

\begin{figure*}[tbp]
	\centering
	\includegraphics[width=\linewidth]{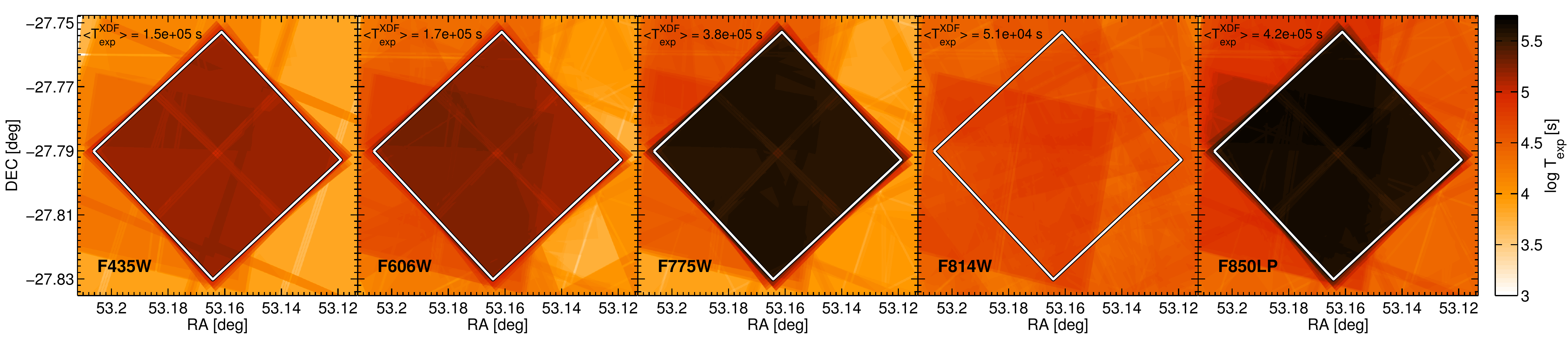}
  \caption{Exposure time maps of the XDF images in the five ACS filters. 
The central diamond shape pointings are from the original HUDF program (Beckwith et al. 2006), 
which took data in all filters except F814W.
  The median exposure time over the HUDF area is indicated on the top, 
ranging from $\sim5\times10^4$ s in F814W to $\sim4\times10^5$ s in F850LP. 
The wider area 
data around the original HUDF imaging have been taken mostly as part of the original GOODS program, and for F814W from CANDELS. In addition, 
a multitude of follow-up programs over the last 10 years have added to this larger area dataset. 
The white outline shows the boundary of the released XDF image to the MAST archive.}
	\label{fig:ExpMapOpt}
\end{figure*}

As would be expected from looking at the distribution of exposures in
Figure \ref{fig:buildup}, the exposure times vary significantly across the
image for the different filters. The total exposure times per filter are
shown in Table \ref{tab:exposures}, while exposure time maps per filter are
presented in Figures \ref{fig:ExpMapOpt} and \ref{fig:ExpMapIR}.

\begin{deluxetable}{lcc}
\tablecaption{XDF exposures \label{tab:exposures}}
\tablewidth{0 pt}
\tablecolumns{3}
\tablehead{\colhead{Filter}	& \colhead{Exposure Time (ks)} & \colhead{\# of Exposures}}
\startdata
\textbf{ACS/WFC} & &\\
F435W & 152.4 & 164\\
F606W & 174.4 & 286\\
F775W & 377.8 & 460\\
F814W & 50.8 & 362\\
F850LP & 421.6 & 700\\[0.05truecm]
\textit{ACS Total} & 1177.0 & 1972\\[0.1truecm]
\textbf{WFC3/IR} & &\\
F105W & 266.7 & 248\\
F125W & 112.5 & 289\\
F140W & 86.7 & 118\\
F160W & 236.1 & 336\\[0.05truecm]
\textit{WFC3 Total} & 702.0 & 991
\enddata
\end{deluxetable}

\section{Data Calibration and Reduction}
\label{sec:datared}
In this section we describe the data reduction process for both the ACS/WFC and WFC3/IR images, and how we aligned these images to the original HUDF data.

Our processing begins by visually inspecting all images included in the \textit{XDF} dataset 
in order to identify any data quality issues. This includes problems due to loss of guiding, 
excessive background or pointing inaccuracies. During the visual inspection we also identify 
images affected by satellite trials and optical ghosts from filter reflections generated by 
bright stars (in ACS/WFC) and updated the data quality array to ensure that these artifacts 
are masked during processing. Any image for which a data quality issue could not be corrected 
was rejected from the dataset and not processed. 

The development of procedures that handled large numbers of images with
arbitrary centering and orientation was very challenging. Key issues that
had to be dealt with included ensuring that the distortion solutions were
correct, that the information used for alignment was correct, that cosmic
rays were handled appropriately so as to ensure that compact sources were
not clipped but the overall CR removal was optimized, and that background
variations were minimized. Combining all these data into a common dataset,
and ensuring that all the data were well-aligned, cleaned of cosmic rays
and artifacts, and were photometrically reliable, was a time-consuming task
that took considerable effort from its inception early in 2012 until
submission to MAST in 2013.

\subsection{Pre-processing ACS/WFC images}
\label{sec:ppacs}

The ACS/WFC channel consists of two $4096\times2048$ pixel detectors at a
scale of $~0\farcs05$ pixel$^{-1}$ providing a $~202''\times202''$ field of
view. All ACS/WFC images used in the \textit{XDF} dataset were processed
through the most recent version of the ACS calibration pipeline
\texttt{calacs} (2012.2). The standard calibration process includes bias
subtraction, dark current correction, bad pixel masking and flat-fielding.
In addition to these calibration processes, images taken after \textit{HST}
Servicing Mission 4 (SM4), have been corrected for Charge Transfer
Efficiency (CTE) degradation (Anderson \& Bedin 2010), bias shift, bias
striping (Grogin et al. 2011), and amplifier crosstalk (Suchkov et al.
2010).
 
\subsection{ACS/WFC image reduction process}
\label{sec:rpacs}

The \textit{XDF} ACS/WFC dataset was processed by the data reduction
pipeline \texttt{APSIS} (Blakeslee, et al. 2003). The reduction
process is quite similar to the process used by the
\texttt{MultiDrizzle} software package (Koekemoer et al. 2002) where
each image is passed through a full drizzle-blot-drizzle cycle. Images
are background subtracted and drizzled onto a common tangential-plane
pixel grid. These images are median stacked and the median stack image
is blotted back to each input image position and used as a template
for cosmic-ray rejection. A cosmic-ray mask is generated for each
image and combined with the data quality array. The final image mosaic
is produced by drizzling all the input images onto a single image mosaic 
and combining them with inverse-variance weight maps that take into account
all noise sources (readout, dark current and background noise).

While \texttt{APSIS} is perfectly adequate for producing science quality
images when combining datasets consisting of tens of images, the
\textit{XDF} dataset, with hundreds of images, required further
processing.

\subsubsection{ACS/WFC image registration}
\label{sec:imreg}

Since the pointing accuracy of \textit{HST} is only good to
within a few arc-seconds, the WCS of each image must be refined in
order to acquire precise registration across all images within the 
ACS/WFC dataset.  \texttt{APSIS} performs the image registration
process using software we developed called \texttt{superalign}
(see Appendix~\ref{sec:super}). Briefly, \texttt{superalign} takes as
input, both source positions from a reference catalog and source
positions from catalogs generated for each image (after rectifying
each catalog to remove the geometric distortion). \texttt{superalign}
then uses these positions to compute accurate shift and rotation
corrections that are applied to each input image.  The input reference
catalog was generated from an image produced by combining the original
HUDF F775W image with an astrometrically calibrated GOODS mosaic to
provide accurate alignment both over and outside the HUDF
area.

\begin{figure}[tbp]
	\centering
	\includegraphics[width=0.85\linewidth]{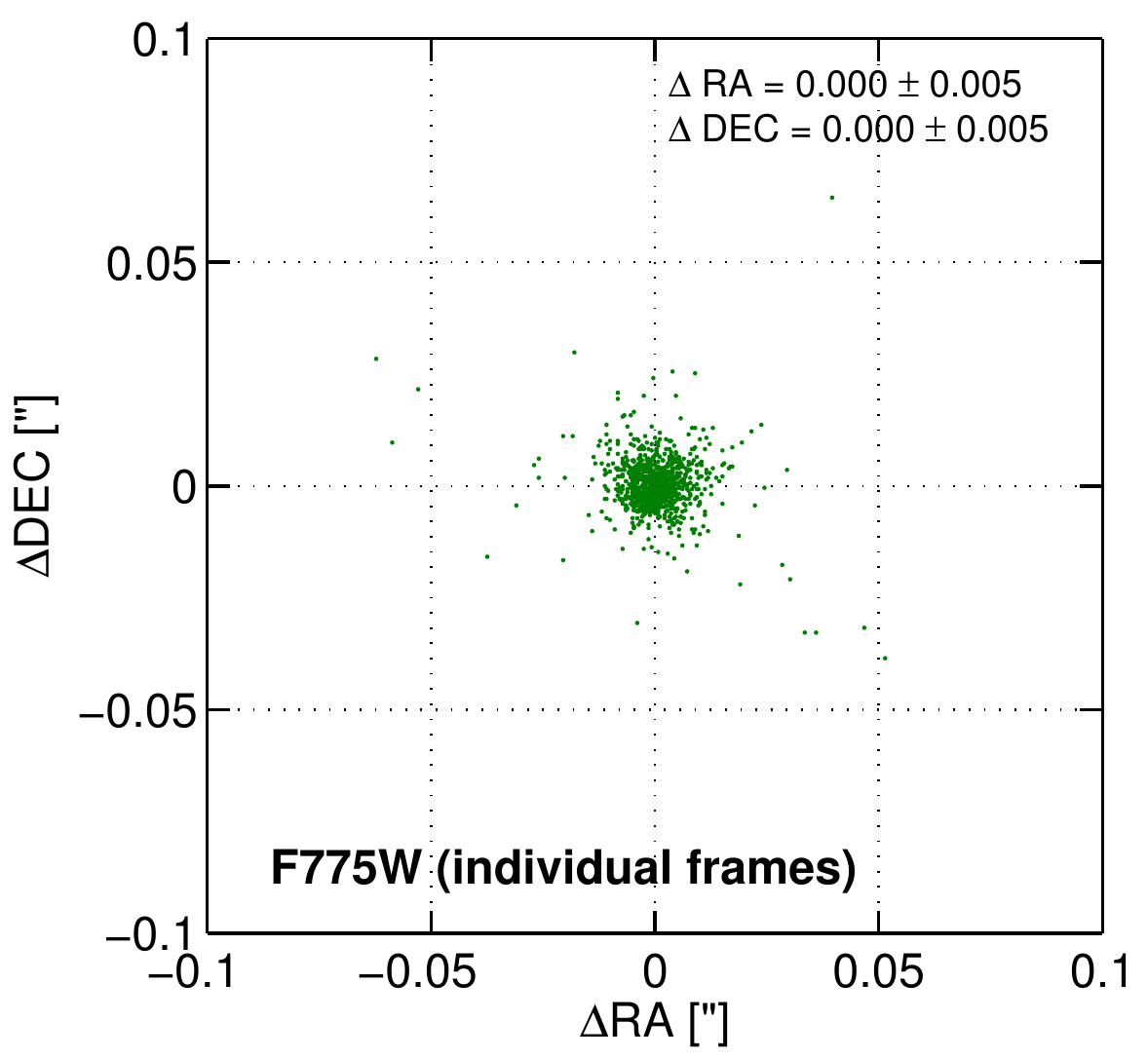}	
  \caption{Relative alignment of individual frames of the XDF F775W data. The
plot shows the offsets of five galaxies and stars in the individual images
relative to the position of these sources in the final XDF image.  There
are 1972 individual frames in the final optical XDF image. This figure shows that
the individual frames align to {\it much better} than a single original ACS pixel
(0\farcs05). The standard deviation in RA and DEC offsets are reported on the
upper right of the plot, indicating that individual frames are aligned to
within 1/10 of a pixel.
}
	\label{fig:AlignmentF775WIndividual}
\end{figure}

To optimize alignment within the ACS/WFC dataset we split the images into
two groups: a deep group with images from the original HUDF observations
that are observed at a single position and two orientations, and a shallow
group, made up of the remaining images, which are observed at a large
variety of positions and orientations.  These two groups are processed
separately with \texttt{superalign} while using the same input reference
catalog. The frame-by-frame alignment within our dataset is excellent as shown in Figure \ref{fig:AlignmentF775WIndividual}.

\subsubsection{ACS/WFC final mosaics}
\label{sec:acsfm}

To create the ACS/WFC mosaics we process the dataset for each filter
through the \texttt{APSIS} pipeline in three passes. The purpose of
the first two passes are to remove any excess background emission on
individual exposures.  This background emission only becomes evident
when stacking a large number of overlapping exposures.  Starting with
the standard calibrated images (\textit{flt} or \textit{flc} files)
acquired from the \textit{HST} archive as input, we create
super-median stacked images for each detector and each filter in the
first pass. The median images are created by halting the pipeline
processing after the blotting process and using the individual blotted
images to aggressively mask sources when combining the stacked
super-median images. These are subsequently subtracted from the
individual images.  This process is repeated in the second pass, but
this time using the subtracted images from the first pass as input and
relaxing the masking threshold. Again, these
super-median images are subtracted from the individual input images,
which are then used in the final pass where the full pipeline is run
to create the final image mosaics. 
These are created with multidrizzle and combine all the input images with the use of inverse-variance weight maps.

Given the large number of input
images used in each filter we elected to use the ``point'' kernel and
a pixfrac of 0 in the final drizzling process.  In making use of this
pixfrac, we remain consistent with the original 2004 release\footnote{http://archive.stsci.edu/pub/hlsp/udf/acs-wfc/} by
\citet{Beckwith06} who use the same pixfrac.

As expected it was not a simple or quick process to generate the final
XDF mosaic.  Processing the full ACS/WFC dataset through our pipeline
involved processing 1.1 terabytes and took 10 days, and so each
processing iteration was a lengthy effort. As problems were found, the
origin of the problems needed to be identified, a fix made, and the
full dataset processed again. Iterating to a final version suitable
for a MAST release required many months of effort.

\begin{figure*}[tbp]
	\centering
	\includegraphics[width=\linewidth]{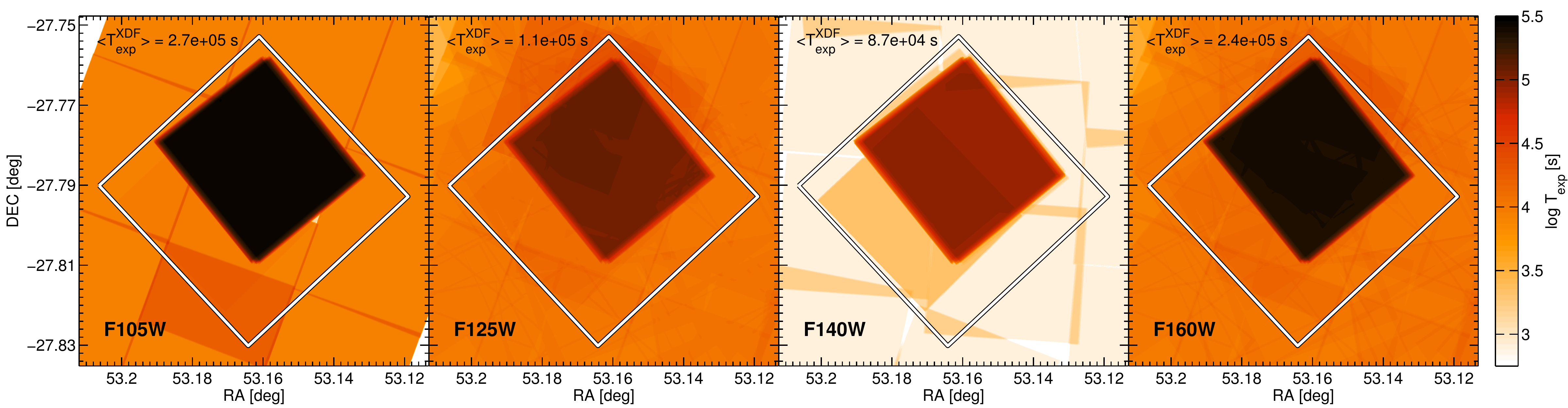}
  \caption{Exposure time maps of the XDF images in the four WFC3/IR filters. 
The vast majority of images in these filters comes from a WFC3/IR pointing
position
that was common to both the HUDF09 and HUDF12 programs. 
  The median exposure time in the deepest part is indicated on the top. The colorbar indicates the range of exposure times.
The total exposures range from $\sim9\times10^4$ s to $\sim3\times10^5$ s. 
The wider area data have been taken mostly as part of the CANDELS Deep program.}
	\label{fig:ExpMapIR}
\end{figure*}

\subsection{Pre-processing WFC3/IR images}
\label{sec:ppir}

Our basic processing of the WFC3/IR images is based on the \texttt{calwf3}
pipeline as outlined below.

The WFC3 IR channel uses a $1024\times1024$ pixel detector with the outer 5
pixels on each side of the detector containing reference bias pixels
yielding a final $1014\times1014$ px image at a scale of
$~0\farcs135\times0\farcs121$ pixel$^{-1}$ and a $~131''\times121''$ field
of view.  All WFC/IR images are obtained in MULTIACCUM mode in which each
image contains two short bias readouts followed by a number of
non-destructive readouts as determined by the NSAMP parameter set in the
Phase II proposal. 

The standard \texttt{calwf3} processing of a WFC3/IR image includes
initializing the data quality array and flagging known bad pixels,
subtracting the mean bias level computed from reference pixels surrounding
the detector for each readout and then subtracting the zeroth (bias) read
from each readout in order to remove any signal from external sources. It
then proceeds by correcting for the non-linear detector response,
subtracting the appropriate dark current reference image in each readout,
computing photometric header keywords and converting the units of the
science and error data arrays to a count rate. This is followed by an
``up-the-ramp'' fitting in order to combine the data from each readout
while identifying and flagging pixels suspected of cosmic-ray hits. The
final step in the processing corrects for pixel-to-pixel and large scale
variations across the detector by dividing by the appropriate flat field image
and then multiplying by the gain of the detector so the final units will be
in electrons per seconds. This step produces the final calibrated image with the
\textit{flt} name extension.

\subsubsection{Variable background correction}
\label{sec:varbc}

While most WFC3/IR \textit{flt} files produced by \texttt{calwf3} can be
used without further calibration, in some cases \texttt{calwf3} can falsely
flag a large fraction pixels as cosmic-rays. For WFC3/IR MULTIACCUM mode
observations, \texttt{calwf3} assumes that accumulating background counts
over the entire observation are linearly increasing. This assumption may
not be the true for all observations, in particular, if the background
varies across an exposure. This can cause \texttt{calwf3} to reject a
significant fraction of the accumulated pixel fluxes, effectively resulting
in a shorter exposure time in affected pixels. The TIME array extension of
the \textit{flt} files can thus be used to determine whether
non-linearities in the background count rate are a problem. 

We tested different criteria, and found that problematic frames can be
identified by an average exposure time of the TIME array that varies by
more than 2\% from the header EXPTIME value. For such images, we introduce
one additional step to the \texttt{calwf3} processing. 

We begin by acquiring the \textit{raw} image and running the
\texttt{calwf3} tasks DQICORR, ZSIGCORR, BLEVCORR, ZOFFCORR, NLINCORR,
DARKCORR, and FLATCORR. We then halt the processing and scale the sky
background in each of the individual MULTIACCUM readout science arrays to
match the average sky count rate across the exposure. This additional
processing step ensures the background count rate to be linear, before
processing the image with the \texttt{calwf3} cosmic-ray rejection task
CRCORR. After background subtraction, the processing is concluded by
running the \texttt{calwf3} task CRCORR, UNITCOR and PHOTCORR, generating a
new cosmic-ray corrected calibrated \textit{flt} image.

\subsubsection{WFC3/IR Persistence masking}
\label{sec:persist}

The WFC3/IR detector can exhibit ghost sources from bright objects imaged
in earlier exposures due to persistence. For the \textit{XDF} WFC3/IR
dataset we therefore excluded all pixels that are significantly affected by
source persistence. To determine which pixels to exclude we utilized the
persistence models generated by the STScI WFC3 Persistence
Project\footnote{\url{http://archive.stsci.edu/prepds/persist/index.html}}.
The project generates a persistence model for each WFC3/IR exposure based
on the time history of previous exposures, incorporating internal
persistence (from exposures within a visit) and external persistence (from
exposures from earlier visits). These are combined to create a model of the
total (internal plus external) persistence flux per pixel. 

We tested different masking criteria, and we found that excluding all
pixels with model persistence flux of 0.2 electrons/s and growing this mask
by 2 pixels, results in clean images that are not significantly affected by
source persistence.

\subsection{WFC3/IR image reduction process}
\label{sec:rpir}

The \textit{XDF} WFC3/IR dataset was processed using tasks provided in the
data reduction pipeline \texttt{WFC3RED}. For a general description of the
\texttt{WFC3RED} pipeline see Magee, Bouwens \& Illingworth 2011. The
\texttt{WFC3RED} pipeline takes as input the pre-processed WFC3/IR
\textit{flt} images and an external reference image used for registration
and performs background subtraction, image registration, creation of
inverse variance weight-maps (for drizzling) and generation of the final distortion-free
image mosaics.

The background in WFC3/IR images can vary dramatically from image to image
depending on the observational conditions and can contain features on
various scales. The \texttt{WFC3RED} pipeline preforms background
subtraction by utilizing a two-pass background model. In each pass, a
median filter is applied to the image and subtracted after aggressively
masking sources. In the first pass a large grid size (165 pixels) is used
in order to remove any large scale gradients. In the second, a smaller grid
size (39 pixels) is used to remove smaller scale background features. In
order to improve the pixel-by-pixel S/N, and correct for any imperfections
in the flats or darks, \texttt{WFC3RED} median stacks all the background
subtracted images in each filter in the dataset to create super-median
background images. These are subsequently subtracted from the individual
images.

\texttt{WFC3RED} performs the image registration process in two steps. The
first registration step is preformed by \texttt{superalign}. While
\texttt{superalign} accurately determines the shift and rotation correction
to be applied to most of the WFC3/IR images we find that for some images a
small correction is needed. 

As a second registration step, \texttt{WFC3RED} therefore uses
\texttt{MultiDrizzle} to create distortion-free image stacks for each
visit in the dataset and then cross-correlates the sources in each
image stack with sources in the external reference image.  Minor
corrections to the shift and rotation are then applied to the WCS of
each image in the visit to obtain an optimal registration.

Specifically, for the alignment of the XDF dataset, we used the original
HUDF F775W image for registration, combined with astrometrically calibrated GOODS mosaics, 
to provide accurate alignment outside of the HUDF area. Thanks to
our two step alignment procedure, the registration of the $XDF$ image to
the HUDF is better than 1/10 of a pixel (i.e., $<0\farcs003$; see section
\ref{sec:HUDFchecks}).


\texttt{WFC3RED} creates the final image mosaics for each filter by running
\texttt{MultiDrizzle} in three passes. For the first pass, each filter
dataset is processed using the full drizzle-blot-drizzle cycle. Each
individual image is drizzled to a separate undistorted image matching the
same size, scale and orientation as the external reference image (i.e., the
HUDF image in the case of the XDF). These images are then stacked into a
single image using the \texttt{MultiDrizzle} ``median''
algorithm\footnote{Note that the standard algorithm in
\texttt{MultiDrizzle} for the stacked image is ``minmed", which works very
well for a small number of input images. However, in the case of the
\textit{XDF}, it can cause central pixels of bright sources to be rejected,
which is why it was not adopted here.}. The stacked median image is then
blotted back to each of the input image (distorted) positions, rescaled by
the exposure time, and used as a clean template for identifying cosmic-rays
and bad pixels. A cosmic-ray/bad-pixel mask is created and the data quality
array is updated for each image. Lastly, a combined drizzle image is
produced for each filter by using the inverse variance weight maps.

In order to remove any excess background emission in the final image
mosaics, \texttt{WFC3RED} runs \texttt{MultiDrizzle} a second time. In this
pass only the blotting back process of \texttt{MultiDrizzle} is run, but
this time using a combined drizzle image from the first pass to create an
image at each position. These images are then used to create a median stack
of images in each filter -- but now masking out the sources apparent in the
combined drizzle image (rather than just those apparent in the individual
images). These super-median images are subtracted from each of the
individual images.

In the last pass only, the final image combination step of
\texttt{MultiDrizzle} is run. A final image mosaic is created for each
filter, using inverse variance weight-maps with a pixfrac\footnote{In drizzling, the pixfrac refers to the size of the footprint of a pixel in units of the input pixel size.} of 0.8, a
``square'' kernel, and  an output scale of $0\farcs06$ pixel$^{-1}$.

The WFC3/IR dataset involved a slightly smaller but still comparable
effort to that involved in processing the ACS/WFC dataset. Nearly 1000
exposures totaling $\sim 0.7$ million seconds in the 4 wide WFC3/IR
filters were processed to form the extremely deep near-infrared
dataset for the XDF. The two primary components to the WFC3/IR data
were from the HUDF09 and the HUDF12 programs, but with additional
contributions from the CDF-S CANDELS dataset.

While the two primary datasets were closely aligned, the CANDELS data
was at different orientations and centers and added to the complexity
of the task for the WFC3/IR data. Processing the WFC3/IR data has many
challenges similar to those of the ACS, but the WFC3/IR data also
bring some unique challenges to the table. The issues that required
particular attention were the mis-flagging of pixels as cosmic rays,
persistence effects and dealing with dramatically varying
backgrounds. Fortunately the smaller WFC3/IR datasets ($\sim 300$
Gbyte) processed more rapidly (typically within a day) and so it was
possible to evaluate the output of a processing run, derive a fix for
a problem and do another run with a reasonable turnaround of several
days.  Nonetheless, the total time to identify the data problems,
refine the software, test and fully process the data, and iterate as
needed, meant that the WFC3/IR data also took many months of effort.

\begin{deluxetable}{lcc}
\tablecaption{XDF Image Depths \label{tab:5SigmaDepths}}
\tablewidth{0 pt}
\tablecolumns{3}
\tablehead{\colhead{Filter} & $5\sigma$ Depth  & $5\sigma$ Depth \\
 & Aperture  & Total  }
\startdata

\multicolumn{3}{c}{\textbf{ACS/WFC}}\\
F435W & 29.8 & 29.6\\
F606W & 30.3 & 30.1\\
F775W & 30.3 & 30.1\\
F814W & 29.1 & 28.9\\
F850LP & 29.4 & 29.2\\
\multicolumn{3}{c}{\textbf{WFC3/IR}}\\
F105W & 30.1 & 29.7\\
F125W & 29.8 & 29.4\\
F140W & 29.8 & 29.3\\
F160W & 29.8 & 29.3 \\

\multicolumn{3}{c}{\textbf{Combined for flat $f_\nu$ Source}} \\
ACS & 30.8 & 30.6 \\
ACS+WFC3 & 31.2 & 30.9 

\enddata
\tablecomments{Depths are measured in circular apertures of 0\farcs35 diameter on the 60 mas XDF images over the part where the WFC3/IR data are deepest (see Figure \ref{fig:ExpMapIR}). The corrections to total magnitudes (last column) are based on the encircled energy tabulated in the ACS and WFC3/IR handbooks.   }
\end{deluxetable}

\begin{figure*}[tbp]
	\centering
	\includegraphics[width=\linewidth]{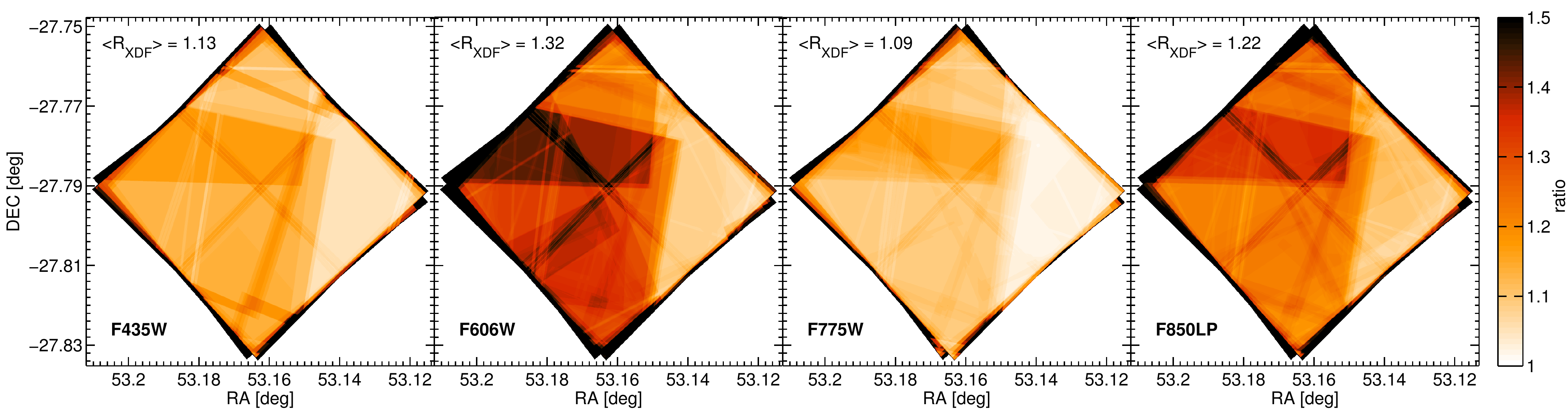}
  \caption{Gain in exposure time of the XDF images compared to the original HUDF data. The 
largest gains are obtained in F606W and F850LP. The largest gains from additional data 
in the follow-up programs were taken in these filters. The gain scale is shown at the right. 
The largest contribution of additional data came from  
our HUDF09 program, which added ACS parallel data that covers the left $\sim75\%$ of the XDF image. 
The median exposure time gains over the XDF image are indicated in the upper left corner of each panel.}
	\label{fig:ExpRatioOpt}
\end{figure*}

\begin{figure}[tbp]
	\centering
	\includegraphics[width=0.9\linewidth]{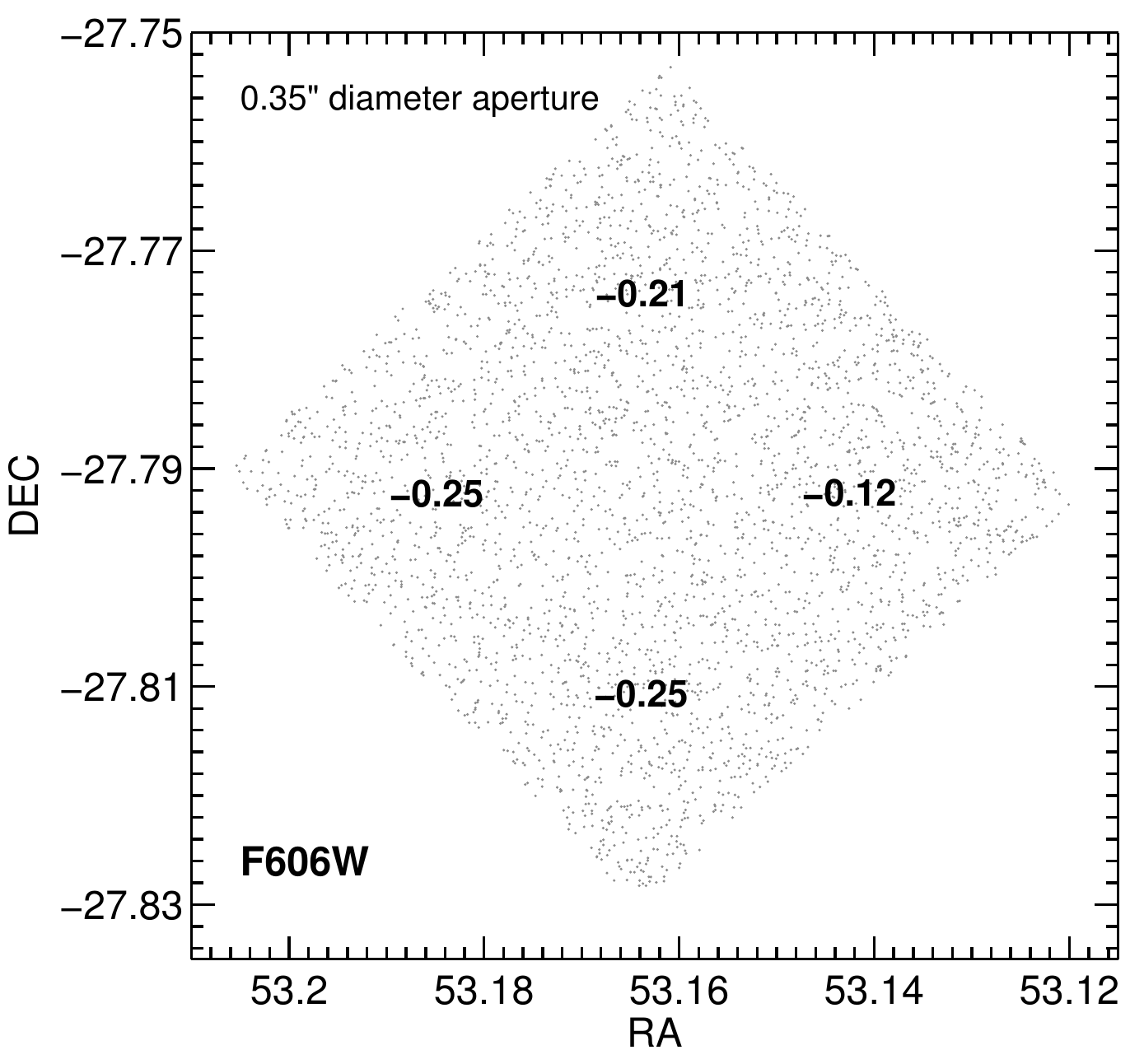}
  \caption{Gains in 5$\sigma$ depth of the XDF image compared to the original HUDF 
image in the F606W filter. The small dots correspond to empty sky positions which were used 
to measure the flux fluctuations in small circular apertures of 0\farcs35 diameter. The gains in depth vary 
between 0.12 mag to 0.25 mag across the image due to the varying gains in exposure time. The average gain in 
F606W is 0.18 mag. These gains are somewhat larger than expected purely based on the 
added exposure time. In particular, our subtraction of a sky image results in markedly smoother background 
than achieved in the original HUDF images leading to a gain in depth of
about 0.1 mag just from the improved background.  }
	\label{fig:DepthGainZ}
\end{figure}

\begin{figure}[tbp]
	\centering
	\includegraphics[width=0.99\linewidth]{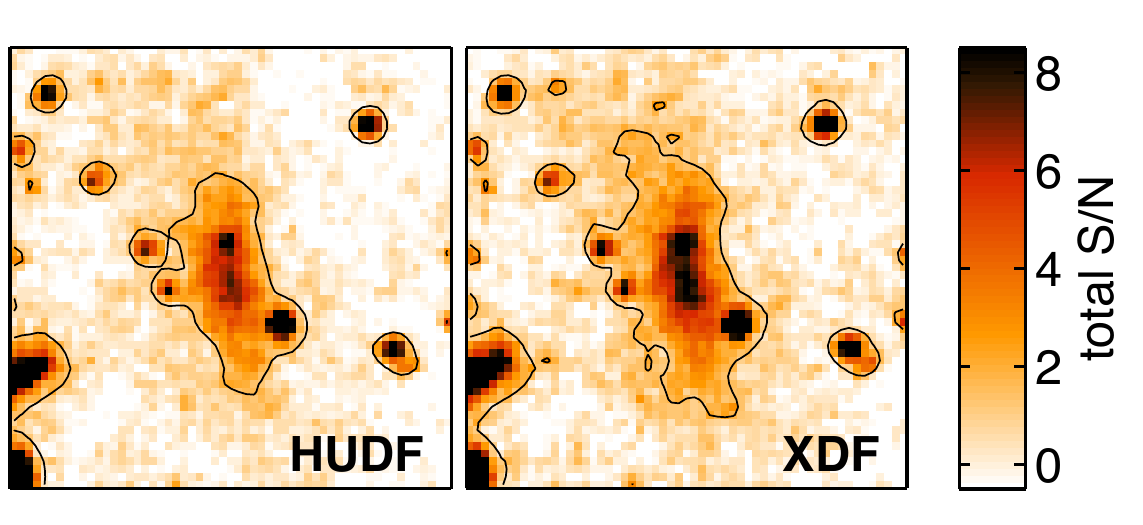}
  \caption{Images of signal-to-noise (not intensity) comparing the improvement of S/N
in the combined ACS optical data set for a small 2\farcs5 region around a
low-surface brightness galaxy in the HUDF/XDF. The pixel-by-pixel S/N was
computed for a pixel size of 0\farcs09 in the total stacked ACS data which
included all images in the F435W, F606W, F775W, and F850LP filters.  This
is shown for the HUDF on the left and for the XDF on the right at the same
color stretch, as indicated by the colorbar on the right that shows the
corresponding S/N.  The black contours indicate where the S/N is larger
than 1.5. Clearly, both the lower surface brightness parts and the core of
this galaxy are detected more significantly in the new XDF image.
   }
	\label{fig:ImprovementFigure}
\end{figure}

\subsection{Data Products}
\label{sec:dataprod}

The \textit{XDF} data products are organized into sets of images by
passband (ACS/WFC F435W, F606W, F775W, F814W \& F850LP; WFC3/IR F105W,
F125W, F140W \& F160W) and image scale. We drizzled the data to two
different scales 0.06$''$/pixel and 0.03$''$/pixel. The ACS benefits
from drizzling to a 0.03$''$/pixel scale because of its smaller native
pixels and the better point-spread function (PSF) at shorter
wavelengths, and so we provide the ACS data at this scale.  These
30mas images are exactly aligned with the original HUDF ACS images,
having the same pixel positions and WCS configuration.

For the WFC3/IR data, a 0.06$''$/pixel scale is more appropriate. We
provide matched ACS and WFC3/IR data at the 60 mas scale. Each 60
milli-arcsecond/pixel image is approximately 5k $\times$ 5k pixels and
each 30 milli-arcsecond/pixel image is approximately 10k $\times$ 10k
pixels. For each filter we provide both the science and inverse
variance weight image.

The full set of \textit{XDF} data products are available through the MAST
High Level Science Products (HLSP) archive at
\url{http://archive.stsci.edu/prepds/xdf/}.

\begin{deluxetable}{lccc}
\tablecaption{$5\sigma$ Depths of XDF compared to HUDF ACS images\label{tab:depth}}
\tablewidth{0 pt}
\tablecolumns{4}
\tablehead{\colhead{Filter} & XDF Depth & XDF Range  & HUDF Depth  }

\startdata
F435W  &  29.72 & 29.69 - 29.79 & 29.60\\
F606W  &  30.20 & 30.14 - 30.27 & 30.02 \\
F775W  & 30.26 & 30.23 - 30.31 & 30.10 \\
F850LP  & 29.43 & 29.41 - 29.46 & 29.23 

\enddata

\tablecomments{The limits correspond to $5\sigma$ variations in the sky flux measured in
a circular aperture of 0\farcs35 diameter on the 30 mas XDF ACS images.  }

\end{deluxetable}

\section{XDF Characteristics and Gains}
\label{sec:gains}

In the following sections, we present some of the characteristics of
the XDF images and compare them to the original HUDF ACS images
\citep{Beckwith06} as well as the previous release of the WFC3/IR data
over the HUDF as part of the HUDF12 program
\citep{Ellis13,Koekemoer12}.

\subsection{Improvement in Depth Relative to HUDF Images}

In Figure \ref{fig:ExpRatioOpt}, we show the gains in exposure time of
the XDF image relative to the original HUDF ACS images. Since many of
the additional data were taken as parallel images (mostly as part of
the HUDF09 program), the added exposure time is highly position
dependent.  In particular, the HUDF09 parallel ACS observations do not
cover the whole XDF image, resulting in maximal gains in exposure time
over the eastern part of the image (left on the figures), and only
small increase in exposure time on the west (right).

In order to quantify the (position-dependent) gain in the depth of the
images, we selected 5000 random empty sky regions of the images and
measured the flux variation in circular apertures of 0\farcs35 diameter.
This was done both in the XDF and in the original HUDF images. The flux
variations are converted into 5$\sigma$ magnitude depths and are reported
in Table \ref{tab:depth}. As can be seen, averaged over the whole field,
the XDF images are $0.12-0.20$ mag deeper than the original HUDF data. 

We quantify the depth variations around the field seen in Figure
\ref{fig:ExpRatioOpt} in Figure \ref{fig:DepthGainZ} for one of the
ACS filters, F606W.  As expected, the eastern part of the image (left)
shows much greater gains in depth than is average for the image.  Due
to the increases in exposure time, the eastern part is deeper by 0.25
mag in that filter compared to the HUDF image.

While these gains may seem small (0.12 mag to 0.25 mag), it is
worthwhile considering that to achieve these gains through a new
imaging program would require between 100 to 240 orbits of new data,
and so the effort expended on maximizing the return from existing
datasets results in valuable and very cost-effective gains.

Note that the minimum gains are actually quite significant, averaging
about 0.1 mag, even in the parts of the images which only got a very
small increment in exposure time. The minimum gain in the redder
filters is quite substantial, reaching 0.13 mag and 0.18 mag, while in
the bluer filters it is 0.09 mag and 0.12 mag. 
This is mainly due to our subtraction of a sky image, which results in
markedly smoother background compared to the original processing of the
HUDF\footnote{A likely cause for the
  non-uniform background in the original HUDF images was due to the
  presence of the ``herringbone'' artifact which affected the bias
  frames used in the original processing \citep[see discussion
    in][]{Oesch07}.} dataset.  It is clear from the original HUDF images that the
background is smoother in the bluer filters than it is in the redder
filters and this is reflected in the gains that we see from the improved
processing.


To exemplify the gains made, we provide a direct comparison in Figure \ref{fig:ImprovementFigure}
between the original HUDF and the XDF of a galaxy with low surface
brightness structure.

\begin{figure}[tbp]
	\centering
	\includegraphics[width=0.75\linewidth]{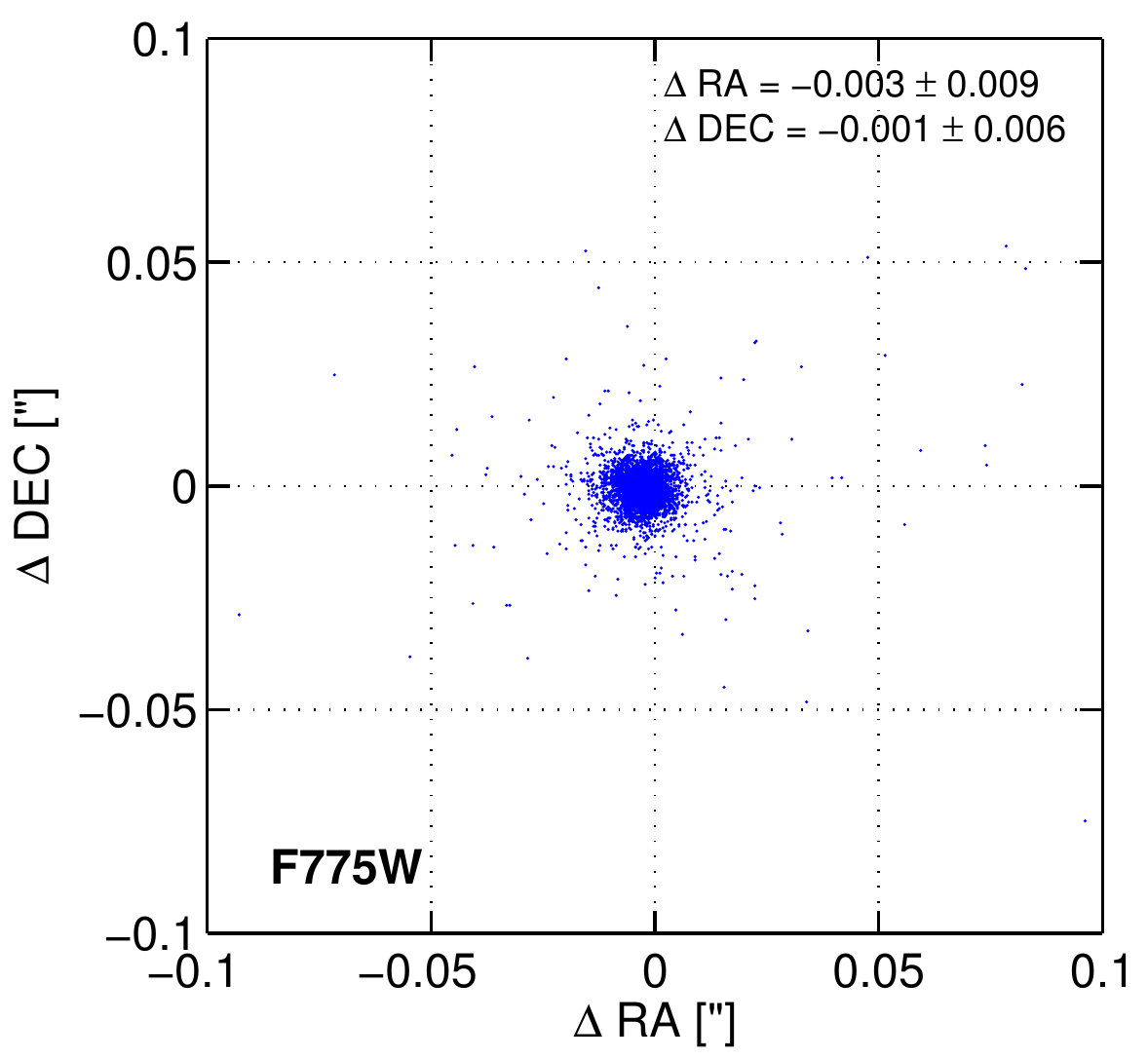}
	\includegraphics[width=0.75\linewidth]{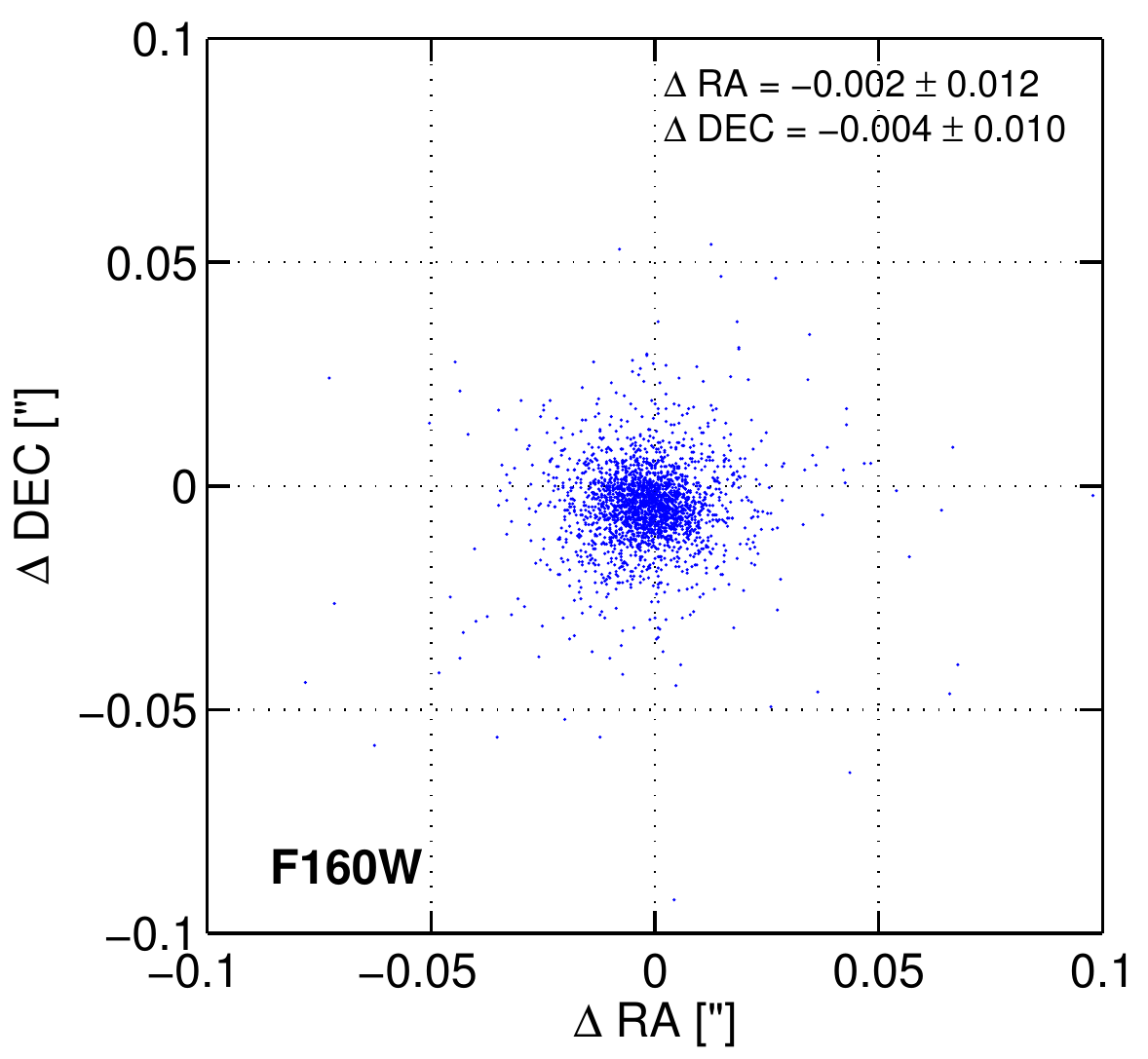}
  \caption{\textit{Top --} Alignment of the XDF F775W data relative to the original HUDF images. 
The plot shows $\sim3500$ galaxies indicating that the alignment is better than 1/10 pixel 
(or $<$3 mas at our px scale of 30mas). The median offsets and the standard 
deviation in RA and DEC are reported on the upper right of the plot. \textit{Bottom -- } Alignment of the XDF F160W image relative to the HUDF12 image release. Given the 60mas pixel scale of the WFC3/IR data, the alignment is again better than 1/10 pixel, or $<4$mas. }
	\label{fig:AlignmentF775W}
\end{figure}

\begin{figure}[tbp]
	\centering
	\includegraphics[width=\linewidth]{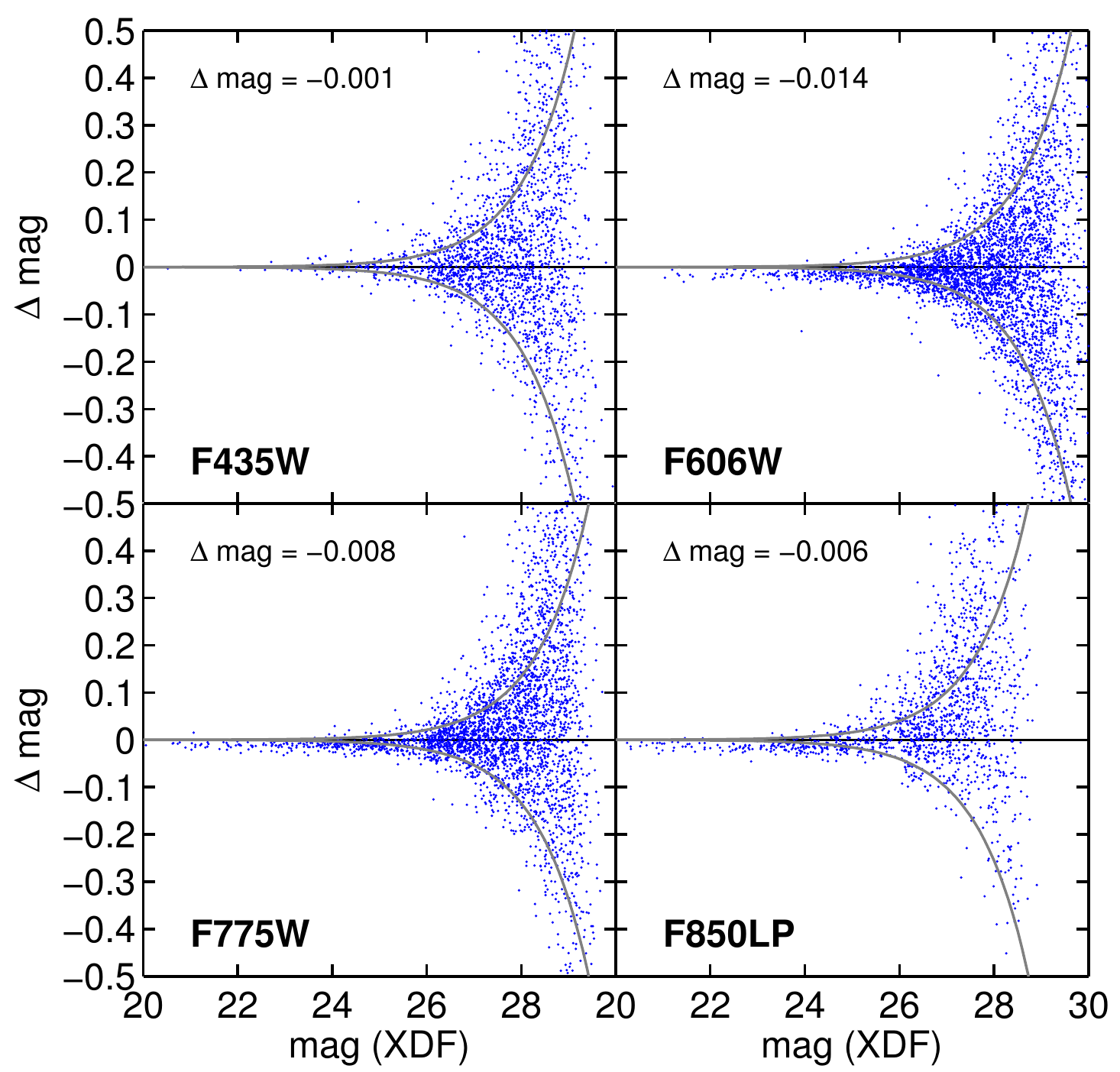}
   \caption{XDF aperture photometry compared to the original HUDF data from identical 
sources in the optical images. Each panel shows a different filter image. Fluxes were measured in circular apertures of 1\arcsec diameter. The photometry is consistent at the 1\% level. 
The gray lines show the expected scatter given the image depth in each filter.}
	\label{fig:PhotometryF775W}
\end{figure}

\begin{figure}[tbp]
	\centering
	\includegraphics[width=\linewidth]{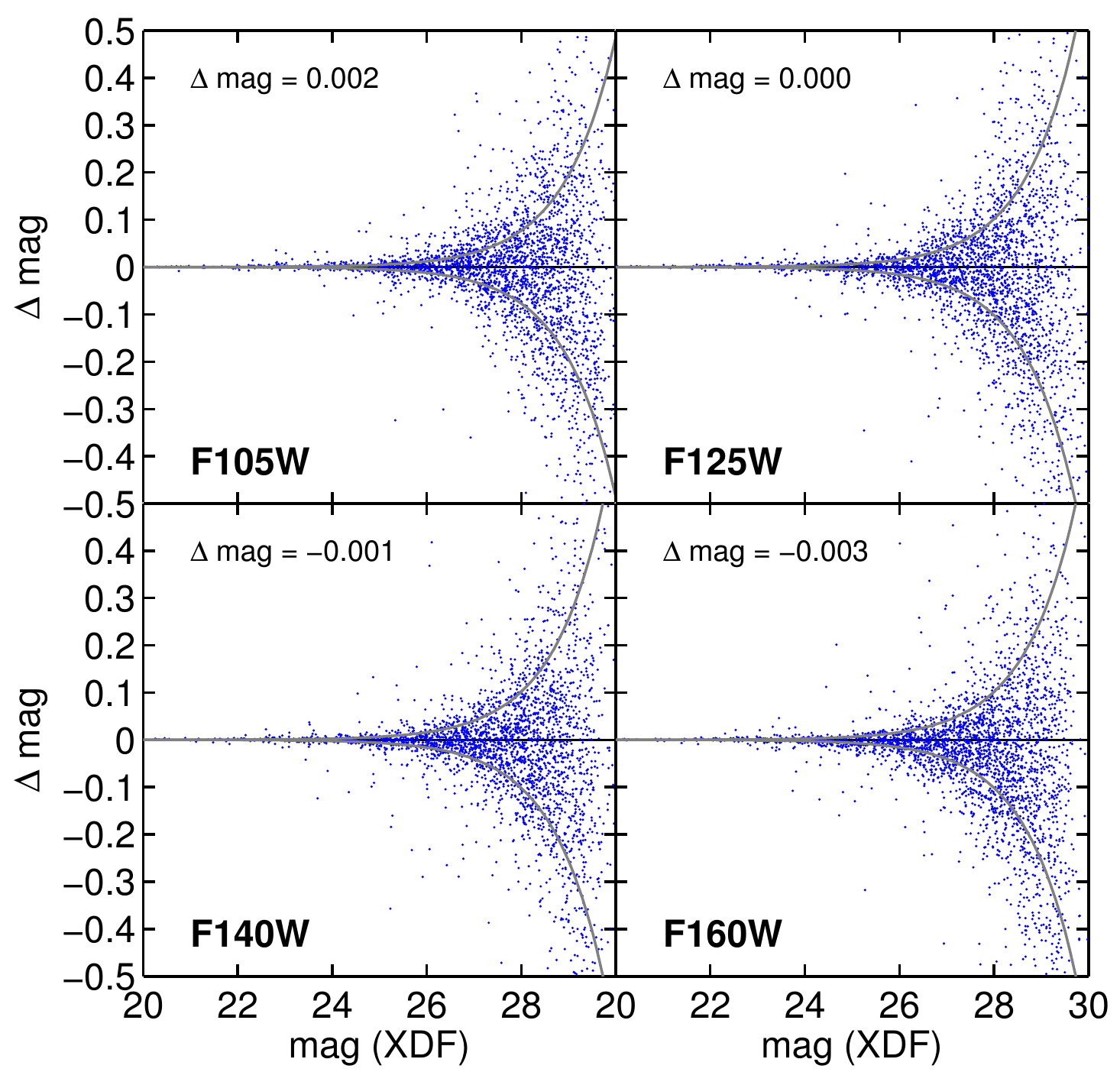}
  \caption{Comparison of the XDF WFC3/IR photometry with the HUDF12 image release. 
Each panel shows the magnitude differences of about 3000 sources measured in circular 
apertures of 1\arcsec diameter. The XDF photometry is consistent with the HUDF12 image 
release at much better than 1\%. The same is true when the XDF photometry is compared 
to the original HUDF09 image over the HUDF. The gray lines show the expected scatter given the magnitude limits in the same apertures.}
	\label{fig:PhotometryIR}
\end{figure}

\subsection{Consistency with Previous HUDF ACS Images}
\label{sec:HUDFchecks}

One important test when creating a new image of the HUDF is consistency
with the previous data. We therefore created independent catalogs of both
datasets for a detailed comparison. In particular, we used the publicly
available images of the HUDF from MAST. 

The weight maps provided by multidrizzle are inverse variance maps. In
order to use these as weight maps for source detection with SExtractor \citep{Bertin96}, we
converted these images to RMS maps using \[ \mathrm{RMS} =
1/\sqrt{\mathrm{WHT}} \] All pixels where $\mathrm{WHT}=0$ are set to 1 in
the RMS-map. 

We then run SExtractor (v2.8.6) to detect sources and measure their
positions (X/Y\_IMAGE) and fluxes in circular apertures (MAG\_APER) of
0\farcs35 diameter.  The source positions are compared in Figure
\ref{fig:AlignmentF775W}, where we show that the alignment is better than
1/10 pixel. The median offsets are $<3$ milli-arcsec, with a standard
deviation of 10 milli-arcsec.

In Figure \ref{fig:PhotometryF775W}, we compare the F775W photometry
in circular apertures between the XDF and the HUDF images (using identical magnitude zeropoints). The
photometry is very consistent, being $\lesssim1\%$ in all cases except
one. The only filter where the flux differences are larger than 1\% is
the F606W image.  The exposure time gain in this filter is highest,
and it is likely that the small changes in the ACS AB-magnitude
zeropoint with time affect this filter most. In our reduction, we have
not accounted for such drifts in the zeropoints over the last 10
years. If photometry to within better than 1\% is required, this
change must be accounted for.  We plan to correct for this effect in a
future release of the XDF images.

\subsection{Consistency with Previous HUDF12 WFC3/IR Images}
\label{sec:HUDF12checks}

There have been two previous releases of WFC3/IR images over the HUDF
field. In particular, after the completion of the HUDF09 program (PI:
Illingworth), our team released a reduction to MAST including the full two
years worth of data totalling 111 orbits of WFC3/IR imaging in the three
filters F105W, F125W and F160W used in the HUDF09 observations. 

Since then, another 128 orbits of WFC3/IR imaging were taken as part of the
HUDF12 program (PI: Ellis). These images included additional F140W imaging
and significantly increased the depth in the F105W filter image. A
combination of WFC3/IR data from the HUDF09 and HUDF12 programs has been
released to MAST by the HUDF12 team. The XDF image release includes our own
reduction of these data together with all additional WFC3/IR data that have
been taken over this field, and we drizzled these to an identical frame as
the ACS imaging data for easy multi-wavelength analyses. 

As we show in Figures \ref{fig:AlignmentF775W} and \ref{fig:PhotometryIR},
the WFC3/IR images of the XDF are in excellent agreement with the HUDF12
(and also the HUDF09) image release, both in terms of photometry and source
positions. Aperture fluxes agree to within $<0.3\%$, and the alignment of
these 60~mas images is again better than 1/10 pixel ($<4$ mas).

\subsection{Number of Sources in the XDF}

The total number of sources in a field, or equivalently, the surface
density of sources, has routinely been used as a gauge to quantify how
faint a field reaches.  The original Hubble
Deep Field North \citep{Williams96} revealed some $\sim$2000-3000
sources within a small 5.7 arcmin$^2$ area (depending on whether the
catalog was based on the $V_{606}$-band image alone or the
$V_{606}+I_{814}$ image).  The Hubble Ultra Deep Field in 2004
\citep{Beckwith06} significantly extended the depth available over the
HDF-North, famously finding some 10,144 sources in a $\sim$11
arcmin$^2$ area or $2\times$ the number of sources per unit area as
was revealed in the original Hubble Deep Field North image.

With the availability of our even deeper XDF exposure over the HUDF
region, we can revisit this question of source counts and the surface
density of sources on the sky to very faint limits.  Weighting the
individual images according to the inverse-variance assuming a flat
$f_{\nu}$ spectrum for sources, we constructed catalogs for the XDF
using first only the optical ACS data over the $\sim$11 arcmin$^2$
full area with the HUDF region and second using the optical+near-IR
data over the $\sim$4.7 arcmin$^2$ footprint originally defined by the
HUDF09 program (see outlines in Figures \ref{fig:ExpMapOpt} and
\ref{fig:ExpMapIR}).

Over the full $\sim$11 arcmin$^2$ area which made up the original HUDF
release, our SExtractor catalogs reveal some 14140 sources on the
coadded $B_{435}+V_{606}+i_{775}+I_{814}+z_{850}$ image with a S/N$>5$
(Kron [1980] apertures, with Kron parameters of 1.6 and 2.5).  The
smaller $\sim4.7$ arcmin$^2$ area over which ultra-deep WFC3/IR
imaging is available contains about 7121 galaxies above the same
$5\sigma$ significance level, using as the detection image the coadded
$B_{435}+V_{606}+i_{775}+I_{814}+z_{850}+Y_{105}+J_{125}+JH_{140}+H_{160}$
image.  


The gains in depth in the optical lead to an $\sim40$\% increase in the
total number of sources over the $\sim11$ arcmin$^2$ HUDF area –- and an
equivalent increase in the source surface density. Over the HUDF09
area where ultra-deep near-IR imaging data are available, we achieve
even greater $\geq$50\% gains in source surface density, as one might expect
given the much greater depth of the collective optical+near-IR data
set.

\begin{figure*}[tbp]
	\centering
	\includegraphics[width=0.8\linewidth]{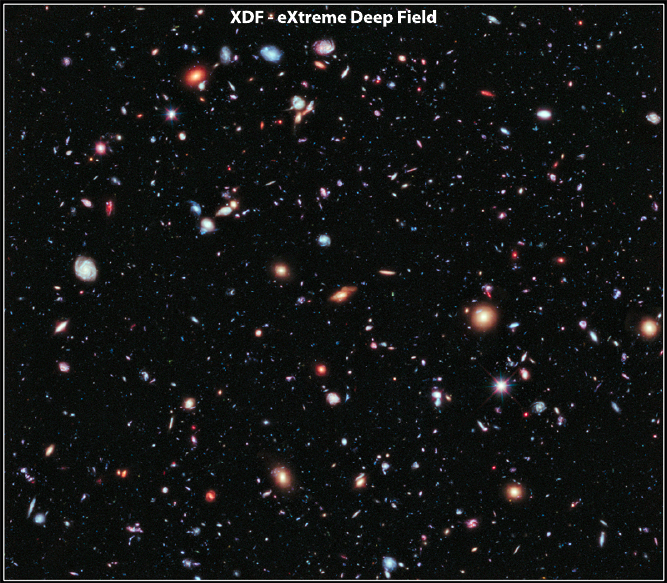}
  \caption{A color image of the deepest region of the XDF, including both the ACS and
WFC3/IR data. The ten years of data on the HUDF region from these cameras
have led to the deepest image ever taken in the optical/near-IR. The point
source $5 \sigma$ depth in a 0\farcs35 diameter aperture in each filter is
typically $\sim$30 AB mag, and is always deeper than 29 AB mag (see Table \ref{tab:5SigmaDepths}).  The combined
depth is $\sim$31.2 AB mag. High resolution versions of this figure can be
found at http://xdf.ucolick.org/xdf.html. }
	\label{fig:colorXDF}
\end{figure*}

\section{Summary} 
\label{sec:summary} 

In 2003 the data HUDF demonstrated the ability of HST and its new ACS
camera to reach beyond 29 AB mag and to push the redshift limit for
galaxies to $z\sim 6$ around 950 Myr after the Big Bang
\citep[e.g.][]{Bouwens04,Bunker04,Yan04}.  Images with the resurrected
NICMOS camera over part of the HUDF showed the potential of near-IR
observations with HST for reaching to even earlier times, $z\sim 7$,
just 800 Myr after the Big Bang \citep{Bouwens04Z,Yan04}, but it was
not until the advent of WFC3/IR in 2009 that this potential was fully
realized.  Images taken as part of the HUDF09 program with WFC3/IR
reached out to $z\sim 10$ (Bouwens et al 2011, but see Ellis et al
2013, Bouwens et al 2013 and Brammer et al 2013 for further
interesting developments from the HUDF12 dataset).  While the data
from these major programs constituted the most extensive programs on
the HUDF, numerous other programs were adding to the available dataset
on the HUDF, but until now, these datasets were not being used to enhance the
HUDF.

The realization in late 2011 that the ten years of observations on the HUDF
from 2002 to 2012 from numerous programs would allow us to push
deeper over the HUDF region led to a program to combine all images from the
ACS and the WFC3/IR into the deepest optical/IR image ever. The dataset
from combining all the available data would provide a resource for numerous
programs on distant galaxies, and would complement the extensive wide-field,
but shallower datasets now available like CANDELS and GOODS.

The goal of the program we undertook was to combine all available ACS and
WFC3/IR data over the HUDF region into a co-aligned dataset that would
result in the deepest ever image. The resulting dataset was named the eXtreme
Deep Field (XDF) to highlight the substantial improvements from taking all
the available data.  The deepest part of the XDF image, in the region
selected initially for the HUDF09 WFC3/IR data, is shown as a color image
in Figure \ref{fig:colorXDF}.

The development of procedures that handled large numbers of images with
arbitrary centering and orientation was very challenging. Key issues that
had to be dealt with included ensuring that the distortion solutions were
correct, that the information used for alignment was correct, that cosmic
rays were handled appropriately so as to ensure that compact images were
not clipped but the overall cosmic-ray removal was optimized, and that background
variations were minimized.  Combining all these data into a common dataset,
and ensuring that all the data were well-aligned, cleaned of cosmic rays
and artifacts, and were photometrically reliable, was a time-consuming task
that took considerable effort from its inception early in 2012 until submission
to MAST in 2013. 

The key steps to realizing the XDF were:
\begin{itemize}
\item Nearly 2000 exposures of ACS data totalling $\sim$1.2 million
  seconds in 5 wide filters were processed to provide the extremely
  deep optical dataset for the XDF. The original HUDF data in 4
  filters dominated the ACS contribution but a large number of
  additional exposures with arbitrary centering and orientation from a
  further 15 HST programs were also incorporated.  These data included
  large numbers of exposures with partial overlap.

\item The resulting ACS image for the XDF added depth through both
additional data and also through improved processing. Much has been learnt about
the ACS data since 2003.  The reprocessing of all the data from the HUDF and the
other 15 programs enabled us to deal with some issues that particularly
affected the background (e.g., post SM4 striping and CTE correction). As a result of the new 
processing approaches and the overall reprocessing the new dataset has smoother
backgrounds.  This provides a typical depth gain of about 0.1 mag. While small, this
is equivalent to adding about 100 orbits of data with the ACS, so the
improvement is an important addition to the HUDF dataset.
The actual depth gains around the ACS field varied from about 0.1 mag to
0.25 mag, corresponding to adding about 100-240 orbits of data to the
original HUDF.

\item The XDF ACS data, for a point source in a $0.35''$ diameter aperture,
reaches to $5 \sigma$ AB mag depths of 29.8 (F435W), 30.3 (F606W, F775W),
29.1 (F814W) and 29.4 (F850LP). The total flux from a point source is 0.2 
mag brighter (i.e., subtract 0.2 mag to give $5 \sigma$ AB mag $total$ 
flux depths).  The ACS dataset has a combined depth for a flat $f_{\nu}$ 
source of 30.8 AB mag  $5 \sigma$ in a $0.35''$ diameter aperture.

\item  The WFC3/IR data for a point source in a $0.35''$ diameter
aperture, reaches to $5 \sigma$ AB mag depths of 30.1 (F105W) and 29.8
(F125W, F140W, F160W). The $5 \sigma$ AB mag depths are $\sim$0.4 mag brighter
for the $total$ flux from a point source.  The WFC3/IR and ACS dataset
together, in the deepest part of the XDF image, has a combined depth for a
flat $f_{\nu}$ source of 31.2 AB mag $5 \sigma$ in a $0.35''$ diameter
aperture.

\item  The processed datasets, ACS and WFC3/IR, were submitted to MAST
and made publicly available on April 5, 2013 (\url{http://archive.stsci.edu/prepds/xdf/}).  
They consist of co-aligned images at 60
mas for both WFC3/IR and the ACS, and a separate set just for the ACS at 30
mas (since the ACS data has higher resolution intrinsically). Both the XDF
ACS and XDF WFC3/IR images are aligned with the original HUDF within a few
mas, or less than $1/{10}$ px. 

\item The WFC3/IR photometry matches within 0.000-0.003 mag (mean
  difference $0.002$ mag) of the HUDF12 release (Koekemoer et al 2013)
  with a $1''$ diameter aperture. The fluxes are also very similar at
  smaller aperture sizes, being 4\% larger in the XDF images.  This
  indicates that the PSF in the new reduction is slightly tighter than
  the HUDF12 reduction.

\item Photometry on the XDF matches the original HUDF within
  0.001-0.014 mag (mean difference $0.007$ mag). The differences are
  small, but measurable and are thought to reflect some aspect of the
  changing zero points with time for the ACS data (resulting from ACS
  detector temperature changes).  While very small, we plan to resolve
  any remaining zeropoint uncertainties for a future release.

\end{itemize}

The remarkable depth of the XDF is unlikely to be exceeded until JWST is
launched, and even then it will not be exceeded in the blue (F435W), 
and possibly not even in the green region of the optical spectrum (F606W), 
until a new space telescope with optical capability is launched.  The XDF 
will remain a cornerstone of the CDF-S and will remain the deepest image 
of the sky for a long time to come. It will be a centerpiece for a wide 
range of studies at high redshift for faint galaxies at redshifts 
around $z\sim2$ out to the limit of Hubble at $z\sim11$.

\acknowledgments{We thank the anonymous referee for very helpful comments on the manuscript. We are grateful to the authors of the MultiDrizzle
  and PyDrizzle software packages Anton Koekemoer and Warren Hack.  
The HUDF and the surrounding observations in the CDF-S from the three
Great Observatories are a unique resource for distant galaxy research.
We thank NASA for funding the Chandra, Hubble and Spitzer missions,
their post-launch operations and science programs, and the numerous
science teams who have contributed to one of the largest
non-proprietary datasets ever taken.   
   Support for this work was provided by NASA through Hubble
  Fellowship grant HF-51278.01. This work has further been supported
  by NASA grant HST-GO-11563.01 and ERC grant HIGHZ \#227749. This work was supported in part by the National Science Foundation under
Grant PHY-1066293 and the Aspen Center for Physics.

The data presented in this paper were obtained from the either the Canadian
Astronomy Data Centre (CADC/NRC/CSA) or the Mikulski Archive for Space
Telescopes (MAST). STScI is operated by the Association of Universities for
Research in Astronomy, Inc., under NASA contract NAS5-26555. Support for
MAST for non-HST data is provided by the NASA Office of Space Science via
grant NNX09AF08G and by other grants and contracts.  }

Facilities: \facility{HST(ACS/WFC3)}.

\appendix
\section{SuperAlign}
\label{sec:super}

\subsection{Overview}

\texttt{superalign} is a short C code we (RJB) developed to determine
the internal shifts and rotations for an arbitrary number of
(overlapping) contiguous images from a set of (distortion free)
catalogs. It requires good initial guesses for the shifts and
rotations (within 2.5 arcsec and 0.5 degrees of the true solution,
respectively), and thus is ideal for use with HST data where these
quantities are only approximately known.  This code was originally
developed for use with the ACS GTO pipeline \texttt{APSIS} and offers
several useful advantages relative to other image registration
packages:

\begin{enumerate}
	\item It does not require that all images be contiguous with a single reference image. This allows one to construct arbitrarily large mosaics out of individual images.
	\item Input catalogs can include substantial ($>80\%$) contamination from cosmic rays.
\end{enumerate}

\subsection{Algorithm}

The algorithm that \texttt{superalign} uses to align large sets of exposures has a tree-like structure:

\begin{enumerate}
	\item First, \texttt{superalign} groups the exposures in terms of those that substantially overlap.
	\item Second, group by group (pointing by pointing), \texttt{superalign} iteratively aligns all the exposures.  As it finds alignments that work between exposures within a group, it constructs an increasingly complete list of the probable stars.  Candidate stars which appear in a statistically significant number of exposures are added to the complete list, while candidate stars which do not are classified as CRs.
	\item After generating a complete list of all the stars from each group, \texttt{superalign} begins building up a mosaic of stars starting at the central group and iteratively accreting nearby groups (to produce an ever larger list of stars). As each group is accreted, the relative position/rotation of all the groups added up to that point (within the ever growing mosaic) is re-optimized to minimize the overall error.
\end{enumerate}

\texttt{superalign} uses a variation on the similar triangle method
(Groth 1986) to determine the alignment between two individual
catalogs. However, instead of attempting to find similar triangles in
both images, \texttt{superalign} looks for similar bi-directional
vectors (object pairs with the same separation and the same relative
orientation). After finding similar vectors, \texttt{superalign} uses
these vectors to determine a candidate transformation from one catalog
onto another. Each transformation is given a score based upon how well
it maps objects from one catalog onto the other.  The transformation
with the highest score is then adopted as the starting point in one
final refinement step, where we perturb the coefficients in this
transformation using a Markov Chain Monte Carlo process in an attempt
to further improve the accuracy of the alignment.


\end{document}